\documentstyle[twocolumn]{article}
\begin{document}

\title{\Large\bf Notes on free fall of a particle and bouncing on a reflecting surface}
\author{{\large Tri Sulistiono}\\
        {\small\em Department of Physics, Institut Teknologi Bandung,
	Jl. Ganesha 10, Bandung, Indonesia 40132 }\\
	{\normalsize (Submitted March 31, 1997)}}
\date{\parbox{140mm}{\small
\hspace*{3mm} Considerable progress has recently been made in controling the motion 
of free atomic particles by means of light pressure exerted by laser radiation. The free 
fall of atoms and bouncing on a reflecting surface made from evanescent wave formed by 
internal reflection of a quasiresonant laser beam at a curved glass surface in the presence 
of homogeneous gravitational 
field has been observed. In this paper we present the energy quantization of this system by 
making use the asymptotic expansion method. It is shown that for large $n$ the levels go 
like $n^{2/3}$ which may be compared with $n^2$ for the infinite square well.\\ 
\\
PACS number(s): 03.65.Bz}}
\maketitle

Considerable progress has recently been made in our understanding of atomic environtment 
behaviour, even which allowed us to control the atom. Atoms have been held near stationary 
\cite{CoPhill}, thrown upwards without heating \cite{Clair,Kasev}, made to produce quantum 
interference after macroscopic path separations \cite{Balyk}, and trapped in quantum wells 
\cite {physto,ajp}. 
Such improvements are mainly accelerated by the invention of a new mechanics called 
wave mechanics or quantum mechanics. Although this new mechanics is different with the 
classical 
mechanics, there are many systems however, showed similar properties both in classical and 
quantum mechanical. One example is the case of simple harmonic oscillator which shows several 
agreement in average energy calculation. As shown in many standard textbooks, 
the average kinetic and potential energies of this system are the same in the two mechanics, 
i.e. $\frac{1}{2}E_0$. One could find that this remains true for the excited levels.

Instead of reviewed the problems on harmonic oscillator, let us now consider the problem of 
free fall of a particle in homogeneous gravitational field and bouncing on an elastically 
reflecting surface. This problem ussualy found in classical, for example, a ball in basketball 
games. This mechanism found possible to exist in the atomic scale. Many experiments already 
performed have shown that multiple bouncing of atoms on a surface can be established 
\cite{prl1,prl2}. The experiment is performed by measuring the average of an ensemble of 
independent atoms. It is known that such systems involved a damping and the (classical) sources 
of damping have been understood \cite{Baly}. The quantum damping and the connection with an 
experiment already performed in the classical regime, however, is studied in \cite{pra}.

In recent paper we focused on the technique of deriving the energy quantization of this system. 
Let us introduce the potential energy of the form
\begin{equation}
V(z) = \left\{
                 \begin{array}{ll}
                      mgz, & (z>0)\\
                      \infty, & (z\leq 0)
                 \end{array}
          \right.
\label{satu}
\end{equation}
which corresponds to the situation of free fall of a particle and bouncing condition, where $m$ 
is mass of the particle, $g$ is the acceleration due to gravity and $z$ is the vertical height 
above the surface. The Schr\"{o}dinger equation for positive $z$ and energy eigenvalue $E$ is 
then 
\begin{equation}
-\frac{\hbar^2}{2m}\frac{d^2\psi}{dz^2}+mgz\psi = E\psi.
\label{dua}
\end{equation}
The boundary conditions of the problem suggest that $\psi=0$ for $z\leq 0$. By introducing the 
caracteristic length of the system,
\begin{equation}
z_0=(\frac{\hbar}{2m^2g})^{1/3}.
\label{tiga}
\end{equation}
it is shown in \cite{pcw} that the solutions can be expressed in term of Airy equation, take 
the form \cite{pra}
\begin{equation}
\phi_n(z)={\rm C}_n{\rm Ai}(\frac{z}{z_0}-\lambda_n).
\label{empat}
\end{equation}
with ${\rm C}_n$'s are normalization constants and $-\lambda_n$ is the $n$th zero of Airy 
function, related to the coresponding energy eigenvalue 
\begin{equation}
E_n=\frac{\hbar^2}{2mz_0^2}\lambda_n.
\label{lima}
\end{equation}

To see the way let us putting $z=\xi z_0+\eta$, where $z_0$ is as described in Eq. 
(\ref{tiga}) 
and $\eta=E/mg$. By doing a little algebra to this identity we can obtain Eq. (\ref{lima}) 
and substitution in Eq. (\ref{dua}) we have then the form 
\begin{equation}
\frac{d^2\psi}{d\xi^2}-\xi\psi=0.
\label{enam}
\end{equation}
which is a form of Bessel's equation. One series solution of the modified Bessel equation 
\begin{equation}
\frac{d^2I}{dx^2}+\frac{1}{x}\frac{dI}{dx}-(1+\frac{p^2}{x^2})I=0.
\end{equation}
is found to be \cite{asymp}
\begin{equation}
I_p(x)=(\frac{1}{2}x)^p\sum_0^\infty\frac{(\frac{1}{4}x^2)^s}{s!(s+p)!}.
\label{tujuh}
\end{equation}
In our problem the solutions are Bessel function of order $1/3$, more commonly known as Airy 
function \cite{pcw} and denoted by ${\rm Ai}(\xi)$ and ${\rm Bi}(\xi)$. The two independent 
solutions are take the form \cite{asymp}
\begin{eqnarray}
{\rm Ai}(\xi)&\!=&\! \frac{1}{3}\xi^{1/2}\{I_{-1/3}(\frac{2}{3}\xi^{3/2})-I_{1/3}
                (\frac{2}{3}\xi^{3/2})\} \nonumber\\
              &\!=&\! \frac{\xi^{1/2}}{\pi\sqrt[]{3}}K_{1/3}(\frac{2}{3}\xi^{3/2}).
\label{ai1}
\end{eqnarray}
\begin{eqnarray}
{\rm Bi}(\xi)&\!=&\! \frac{1}{3}\xi^{1/2}\{I_{-1/3}(\frac{2}{3}\xi^{3/2})+I_{1/3}
                (\frac{2}{3}\xi^{3/2})\} \nonumber\\
              &\!=&\! \frac{2\xi^{1/2}}{\pi\sqrt[]{3}}{\cal K}_{1/3}(\frac{2}{3}\xi^{3/2}).
\label{bi1}
\end{eqnarray}
We can expand the solutions into the form 
\begin{equation}
{\rm Ai}(\xi)=c_1 f(\xi) - c_2 g(\xi).
\label{ai2}
\end{equation}
\begin{equation}
{\rm Bi}(\xi)=\sqrt[]{3}\; [c_1 f(\xi) + c_2 g(\xi)].
\label{bi2}
\end{equation}
where
\begin{equation}
f(\xi)=\sum_{k=0}^{\infty} 3^k\bigl(\frac{1}{3}\bigr)_k \frac{\xi^{3k}}{(3k)!}.
\label{fchi}
\end{equation}
\begin{equation}
g(\xi)=\sum_{k=0}^{\infty} 3^k\bigl(\frac{2}{3}\bigr)_k \frac{\xi^{3k+1}}{(3k+1)!}.
\label{gchi}
\end{equation}
\begin{eqnarray}
3^k\bigl(\alpha +\frac{1}{3}\bigr)_k=(3\alpha+1)(3\alpha+4)\ldots (3\alpha+3k-2),\nonumber \\
{\rm for which }\;\;\;\; k=1,2,3,\ldots.
\end{eqnarray}
with $(\alpha+ 1/3)_0 =1$, $c_1=0.355028053887817$ and $c_2=0.258819403792807$ \cite{microsoft}.

Once again numerical techniques must be used, but it is interesting to use the asymptotic 
expansion which would take the form
\begin{equation}
{\rm Ai}(\xi)\sim\frac{1}{2}\pi^{-1/2}\xi^{-1/4}\exp(-\varsigma)\sum_{k=0}^{\infty}
              (-1)^k c_k \varsigma^{-k}.
\label{ai3}
\end{equation}
\begin{equation}
{\rm Bi}(\xi)\sim \pi^{-1/2}\xi^{-1/4}\exp(-\varsigma)\sum_{k=0}^{\infty} c_k \varsigma^{-k}.
\label{bi3}
\end{equation}
where
\begin{equation}
\varsigma=\frac{2}{3}\xi^{3/2}.
\label{ss}
\end{equation}
\begin{equation}
c_k=\frac{\Gamma (3k+0.5)}{54^k k! \Gamma (k+0.5)}
\label{gama}
\end{equation}
of these two functions only ${\rm Ai}(\xi)$ are acceptable as wave function because 
${\rm Bi}(\xi)$ cannot be normalized. Thus the solutions described by Eq. (\ref{empat}) holds.
We have then
\begin{equation}
\phi(z)\propto {\rm Ai}(\xi)={\rm Ai}(\frac{z}{z_0} -\lambda_n),\;\;\; 
                              \lambda_n=\frac{\eta}{z_0}.
\label{hasil}
\end{equation}
with $z_0\lambda_n$ represents the $n$th classical turning point \cite{pra}, also described by 
$z_n$.

Applying the boundary condition $\phi=0$ at $z=0$ then requires ${\rm Ai}(-\lambda_n) =0$, and
the energy levels are given by the roots of this equation which is strictly accurate  only as 
$n \rightarrow \infty$. This function is an oscillatory function as described in \cite{pcw} and 
for large $E$ we may employ the asymptotic form
\begin{eqnarray}
{\rm Ai}(-x)\sim \pi^{-1/2}x^{-1/4} [\sin(\varsigma +\frac{\pi}{4})\sum_{k=0}^{\infty}
                (-1)^k c_{2k} \varsigma^{-2k} \nonumber\\
                + \cos(\varsigma +\frac{\pi}{4})\sum_{k=0}^{\infty}
                (-1)^k c_{2k+1} \varsigma^{-2k-1}].
\label{airy}
\end{eqnarray}
where $\varsigma$ and $c_k,\; k=0,1,2,\ldots$ are defined in equation (\ref{ss}) and 
(\ref{gama}) respectively.

Once again by applying the boundary condition we can reject the cosinus term in Eq. 
(\ref{airy}) leaving only the sinus term. Thus we have then
\begin{equation}
{\rm Ai}(-\lambda_n)\sim \pi^{-1/2}\lambda_n^{-1/4} \sin (\frac{2}{3} \lambda_n^{3/2} +
       \frac{\pi}{4}).
\label{sin1}
\end{equation}
and obtain $\sin[(2/3)\lambda_n^{3/2}+\pi/4)]\approx 0$, which has solutions
\begin{equation}
\lambda_n=[\frac{3}{2}\pi(n-\frac{1}{4})]^{2/3}.
\label{lamda}
\end{equation}
substituting $\lambda_n=\eta /z_0$ where $\eta=E/mg$ and $z_0=(\hbar^2/2m^2g)^{1/3}$ we have 
then
\begin{equation}
E_n=[\frac{9}{8} \pi^2(n-\frac{1}{4})^2 m\hbar^2g^2]^{1/3}.
\label{eigen}
\end{equation}
This expression, though not quite exact, is always accurate to better than 1\% \cite{pcw}.

It can be shown that for large $n$ the levels go like $n^{2/3}$ which may be compared with 
$n^2$ for the infinite square well \cite{physto}. This is within the result of \cite{nieto} 
that the spectrum can rise no faster than $n^2$ in the nonrelativistic case. The closer 
spacing is connected with the shallower rate of 
climb of the potential. Further investigation on this system gives much more 
understanding about the atomic behaviour.

An experimental demonstration of this problem in the atomic scale presented in \cite{prl2} 
of which multiple bouncing of cesium atoms on a reflecting surface was observed. The 
reflecting surface was made from evanescent wave formed by internal reflection of a 
quasiresonant laser beam at a curved glass surface. A cold cloud of cesium atoms was dropped 
onto the mirror and observed to rebound more than 8 times. The earlier demonstration of such 
reflection with mirror of a reflectivity close to 100\% has been achieved and the 
quantum-state selective reflection of atoms is observed \cite{prl1}.

Following this result was a proposed scheme to create an atomic cavity on the basis of 
reflection of atoms from a laser field \cite{Baly} where the main parameters of cavity such as 
maximum and  minimum atomic velocity, cavity stability, scheme of atomic injection, maximum 
atom density were defined. Later the quantum damping for such system studied in \cite{pra}. 
All the studies can be regarded as a first step towards an interferometer of Fabry-P\'{e}rot 
type for atomic de Broglie waves \cite{Baly,Wallis}.\\

The author would like to thank M. M. Nieto for comment on the previous edition of this note. 


\vspace*{6mm}


\begin{thebibliography}{99}
\bibitem{CoPhill} For a review, see e.g., C. Cohen-Tannoudji, and W. Phillips, 
        Phys. Today {\bf 43}, No. 10, 35 (1990).
\bibitem{Clair} A. Clairon, C. Solomon, S. Guellati, and W. D. Phillips, Europhys. Lett. 
        {\bf 16}, 165 (1991).
\bibitem{Kasev} M. Kasevitch, D. S. Weiss, E. Riis, K. Moler, S. Kasapi, and S. Chu, 
        Phys. Rev. Lett. {\bf 66}, 2297 (1991).
\bibitem{Balyk} V. Balykin, and P. Meystre, Appl. Phys. B {\bf 54}, 319 (1992). Special issue 
        on Optics and Interferometry with Atoms, edited by J. Mlynek.
\bibitem{physto} For a notes and review, see e.g., G. P. Collins, Phys. Today {\bf 46}, No. 6,
 17 (1993), 
\bibitem{ajp} For notes and review on gravitational well, see e.g., P. W. Langhoff, 
Am. J. Phys. {\bf 39}, 954 (1971).
\newpage
\bibitem{prl1}  V. I. Balykin, V. S. Lethokov, Yu. B. Ouchinikov, A. I. Sidorov, 
        Phys. Rev. Lett. {\bf 60}, 2137 (1988).        
\bibitem{prl2} C. G. Aminoff, A. M. Steane, P. Bouyer, P. Desbiolles, J. Dalibard, and C. 
        Cohen-Tannoudji, Phys. Rev. Lett. {\bf 71}, 3083 (1993).
\bibitem{Baly} V. I. Balykin, and V. S. Lethokov, Appl. Phys. B {\bf 48}, 517 (1989).
\bibitem{pra} R. Onofrio, and L. Viola, Phys. Rev. A {\bf 53}, 3773 (1996).
\bibitem{pcw} P. C. W. Davis, and D. S. Betts, {\it Quantum Mechanics}, 2nd ed., London, 
        Chapman \& Hall, 1994.
\bibitem{asymp} R. B. Dingle, {\it Asymptotic Expansions: Their Derivation and Interpretation}, 
        London, Acad. Press Inc., 1973.
\bibitem{microsoft} Microsoft Corp., {\it Microsoft Fortran Library}, user guide, 1989. 
\bibitem{nieto} M. M. Nieto and L. M. Simmons, Jr., Am. J. Phys. {\bf 47}, 634 (1979).
\bibitem{Wallis}    H. J. Wallis, J. Dalibard, and C. Cohen-Tannoudji, Appl. Phys B {\bf 54}, 
        407 (1992).
\end{thebibliography}
\end{document}